\documentclass[a4paper]{article}
\usepackage[colorinlistoftodos]{todonotes}
\usepackage[utf8]{inputenc}
\usepackage{a4wide}
\usepackage{amsmath,amssymb,hyperref}
\usepackage{bm,bbm,tikz}
\usepackage[english]{babel}
\usepackage{caption,subcaption}
\definecolor{vertexcol}{RGB}{102,194,165}
\definecolor{vertexbascol}{RGB}{153,153,153}
\definecolor{edgetracecol}{RGB}{141,160,203}

\newcommand{\expec}{\mathbb{E}}
\newcommand{\Exp}[1]{\expec\left[#1\right]}
\newcommand{\prob}{\mathbb{P}}
\newcommand{\Prob}[1]{\prob\left(#1\right)}

\title{Contact tracing in configuration models}
\author{Ivan Kryven$^{1}$, Clara Stegehuis$^2$\\
\footnotesize{$^1$ Mathematical Institute and the Centre for Complex Systems Studies, Utrecht University, the Netherlands}\\
\footnotesize{$^2$ Department of Electrical Engineering, Mathematics and Computer Science, Twente University, the Netherlands},
}

\date{\today}

\usepackage[numbers]{natbib}
\usepackage{graphicx}

\begin{document}

\maketitle
     

\begin{abstract}
Quarantining and contact tracing are popular ad hoc practices for mitigating epidemic outbreaks. However, few mathematical theories are currently available to asses the role of a network in the effectiveness of these practices.  In this paper, we study how the final size of an epidemic is influenced by the procedure that combines contact tracing and quarantining on a network null model: the configuration model. 
 Namely, we suppose that infected vertices may self-quarantine and trace their infector with a given success probability.
  A traced infector is, in turn, less likely to infect others. We show that the effectiveness of such tracing process strongly depends on the network structure. In contrast to previous findings, the tracing
  procedure is not necessarily more effective on networks with heterogeneous degrees. We also show that network clustering influences the effectiveness of the tracing process in a non-trivial way: depending on the infectiousness parameter, contact tracing on clustered networks may either be more, or less efficient than on network without clustering.
\end{abstract}

\section{Introduction}
Contact tracing is a frequently used method to control epidemic outbreaks. In this method, individuals who show symptoms of a disease, report themselves and identify their recent contacts which are then tested for the disease. If a contact tests positive, they is being isolated to prevent further spreading of the disease. In this way, an epidemic may be contained in its early stages. 

 The effect of contact tracing has mathematically been investigated by extending compartmental models, such as the SIR model, with an additional rule that infected individuals may be detected and removed with some rate that represents a tracing process~\cite{blum2010, clemencon2008,kretzschmar2020}, or by other differential equation approaches~\cite{fraser2004,ferretti2020}. However, such compartmental models simplify the structure of contact networks by representing it with one numerical parameter.
Complex networks on the other hand may have nontrivial structure, featuring  heavy tailed degree distributions, clustering, and other phenomena. 
For example, the contact network of the HIV/AIDS epidemic in Cuba was found to be well-approximated by a power-law degree distribution~\cite{clemencon2015}, so that the proportion of vertices with $k$ neighbors scale as $k^{-\tau}$. Such degree distributions feature a large variability of node degrees, with vertices of large degrees (also called hubs) being present along with large number of small degree nodes. We will refer to this phenomenon as \emph{degree-heterogeneity}. Furthermore, power-law degree distributions where shown to cause important epidemiologic properties, such as vanishing epidemic thresholds~\cite{pastor2001,boguna2004}, strong finite-size effects~\cite{pastorsatorras2002}, and novel universality classes for critical exponents~\cite{dhara2019}.

 A recent simulation study suggested that contact tracing is more effective on networks with high degree-heterogeneity~\cite{kojaku2020}. Intuitively, high-degree vertices infect more others than low-degree vertices, so that they are also more likely to be traced. Furthermore, quarantining high-degree vertices has a larger effect on the spreading of epidemics than quarantining low-degree vertices. Thus, on these types of networks, contact tracing is expected to be more effective than is predicted in the standard SIR-models due to degree-heterogeneity.
In~\cite{bianconi2020}, this expectation was made more formal by showing that the tracing process becomes more effective when high-degree vertices are likely to install contact tracing apps. 
 
 While approaches in~\cite{kojaku2020,bianconi2020} rely on networks being locally tree-like, many real-world networks violate this condition and feature clustering: they contain a high density of triangles. Simulations suggest that network clustering has a strong positive impact on the effectiveness of the contact tracing process in homogeneous networks~\cite{kiss2005}. In general epidemics, clustering can either speed up, or slow down the spread of an epidemic process~\cite{stegehuis2016}. 
 
 In this paper, we quantify the network effect on the effectiveness of contact tracing, by mapping it to a combination of bond- and site percolation models. We show that the extent to which contact tracing reduces the number of infections highly depends on the exact choice of tracing model. We show that when the tracing process is not immediate, but takes a nonzero amount of time, this drastically affects the outbreak size. 
 We then investigate the effect of degree-heterogeneity and clustering on the effect of contact tracing on the final outbreak size using percolation models and find that clustering can either increase or decrease the effectiveness of tracing processes, depending on the infectiousness of the epidemic. This shows that the interplay between the underlying network structure and the exact choice of tracing process is delicate, and important to take into account.
 
 We first describe the network model and define the tracing process in Section~\ref{sec:model}. Then we show the relation between the success probability of tracing and the characteristic time of the the tracing process. Section~\ref{sec:gfanalysis} analyzes the final outbreak size of our epidemic model with a generating function approach. We then study the effect of inducing clustering in the network in Section~\ref{sec:clustering}.
 
 \section{Network and tracing model}\label{sec:model}
 In this paper, we assume that the underlying network is given by the configuration model, a network model that can generate networks with any prescribed degree distribution $(q_k)_{k\geq 1}$~\cite{bollobas1980}. In the configuration model, every vertex of degree $k$ is equipped with $k$ half-edges, which are paired uniformly at random.
 We assume that the disease spreads on this network as a bond percolation process: it removes each edge independently with probability $1-\pi$. While this is very simple variant of an epidemic process, the final size of a SIR epidemic with constant recovery duration can be identified as the size of the largest connected component after bond percolation~\cite{durrett2006}. 
 In this setting, the effective basic reproductive number $R_0$, or the average number of vertices infected by one infected vertex, is given by $R_0=\pi \Exp{D(D-1)}/\Exp{D}$, where $D$ denotes the degree of a uniformly chosen vertex~\cite{may2001}.

\begin{figure}
    \centering
    \begin{subfigure}[b]{0.3\linewidth}
    \begin{tikzpicture}
    \tikzstyle{vertexbas}=[circle,fill=vertexbascol,minimum size=10pt,inner sep=0pt]
\tikzstyle{vertex} = [vertexbas, fill=vertexcol]
\tikzstyle{edge} = [draw,thick,->]
\node[vertexbas] (a) at (0,2) {};
\node[vertex] (b) at (1.5,1) {};
\node[vertexbas] (c) at (0.5,1) {};
\node[vertexbas] (d) at (-0.5,1) {};
\node[vertex] (e) at (-1.5,1) {};
\path[edge] (a)-- (b);
\path[edge] (a)-- (c);
\path[edge] (a)-- (d);
\path[edge] (a)-- (e);
\node[vertex] (b1) at (1.75,0) {};
\node[vertexbas] (b2) at (1.25,0) {};
\node[vertexbas] (c1) at (0.75,0) {};
\node[vertex] (c2) at (0.25,0) {};
\node[vertexbas] (e1) at (-1.5,0) {};
\path[edge] (b)-- (b1);
\path[edge] (b)-- (b2);
\path[edge] (c)-- (c1);
\path[edge] (c)-- (c2);
\path[edge] (e)-- (e1);
    \end{tikzpicture}
    \caption{}
    \end{subfigure}
    \hspace{0.3cm}
        \begin{subfigure}[b]{0.3\linewidth}
    \begin{tikzpicture}
    \tikzstyle{vertexbas}=[circle,fill=vertexbascol,minimum size=10pt,inner sep=0pt]
\tikzstyle{vertex} = [vertexbas, fill=vertexcol]
\tikzstyle{edge} = [draw,thick,->]
\node[vertexbas] (a) at (0,2) {};
\node[vertex] (b) at (1.5,1) {};
\node[vertexbas] (c) at (0.5,1) {};
\node[vertexbas] (d) at (-0.5,1) {};
\node[vertex] (e) at (-1.5,1) {};
\path[edge] (a)-- (b);
\path[edge] (a)-- (c);
\path[edge] (a)-- (d);
\path[edge] (a)-- (e);
\node[vertex] (b1) at (1.75,0) {};
\node[vertexbas] (b2) at (1.25,0) {};
\node[vertexbas] (c1) at (0.75,0) {};
\node[vertex] (c2) at (0.25,0) {};
\node[vertexbas] (e1) at (-1.5,0) {};
\path[edge] (c)-- (c1);
\path[edge] (c)-- (c2);
\path[draw,very thick,edgetracecol,->] (-1.4,1.25)--(-0.2,2);
    \end{tikzpicture}
    \caption{}
    \end{subfigure}
        \hspace{0.3cm}
        \begin{subfigure}[b]{0.3\linewidth}
    \begin{tikzpicture}
    \tikzstyle{vertexbas}=[circle,fill=vertexbascol,minimum size=10pt,inner sep=0pt]
\tikzstyle{vertex} = [vertexbas, fill=vertexcol]
\tikzstyle{edge} = [draw,thick,->]
\node[vertexbas] (a) at (0,2) {};
\node[vertex] (b) at (1.5,1) {};
\node[vertexbas] (c) at (0.5,1) {};
\node[vertexbas] (d) at (-0.5,1) {};
\node[vertex] (e) at (-1.5,1) {};
\path[edge] (a)-- (b);
\path[edge] (a)-- (c);
\path[edge] (a)-- (e);
\node[vertex] (b1) at (1.75,0) {};
\node[vertexbas] (b2) at (1.25,0) {};
\node[vertexbas] (c1) at (0.75,0) {};
\node[vertex] (c2) at (0.25,0) {};
\node[vertexbas] (e1) at (-1.5,0) {};
\path[edge] (c)-- (c1);
\path[edge] (c)-- (c2);
\path[draw,very thick,edgetracecol,->] (-1.4,1.25)--(-0.2,2);
    \end{tikzpicture}
    \caption{}
    \end{subfigure}
    \caption{The tracing process illustrated. (a) shows the infection tree. Every infected vertex self-reports and quarantines with probability $1-p_s$ (green vertices). (b) After quarantining, a vertex loses all offspring. Furthermore, every self-reporting vertex traces its parent with probability $p_t$ (blue arrows). (c) When a parent is traced, all its infectious contacts are removed with probability $1-\delta$.}
    \label{fig:tracingprocess}
\end{figure}
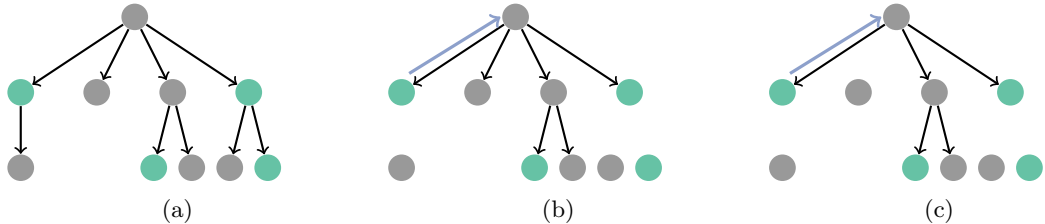

We investigate the effect of a tracing process illustrated in Figure~\ref{fig:tracingprocess} on the final size of the epidemic. In this tracing model, every infected vertex `reports' its infection independently with probability $1-p_s$. After reporting, a vertex quarantines, so that it is unable to infect other vertices, as shown in Figure~\ref{fig:tracingprocess}(b). Furthermore, a vertex that `reports' itself as infectious lists its recent contacts and, with success probability $p_t,$ the infector of the reporting vertex is identified in this list. In this case, we say that the infector vertex was `traced'. 

After a vertex is `traced', it quarantines, so that it is unable to infect other vertices. However, the traced vertex may already have infected other vertices before it was traced. We therefore model such secondary quarantining of the traced vertices by removing each edge incident to a traced vertex with probability $1-\delta$. That is, the tracing process is modelled as an extra layer of bond percolation, see Figure~\ref{fig:tracingprocess}(c).


\subsection{Immediate or delayed tracing: the impact of $\delta$}\label{sec:delta}
 The probability that the connection to a vertex is removed when its parent is traced, $1-\delta$, depends on the parameters $p_s$ and $ p_t$. Here we show how $\delta$ relates these parameters under two assumptions on the tracing process: immediate and delayed tracing, and discuss the impact of these assumptions on the effectiveness of the tracing process.
 
\subsection{Immediate tracing}
 We first assume that the tracing process is immediate: once a vertex self-reports, it immediately traces its parent with probability $p_t$. If successful, the traced vertex immediately quarantines and cannot infect other vertices anymore. We now show that this assumption leads to a degree-dependent version of $\delta$: $\delta_k$.
 
Consider an outcome of the infection process as a tree composed of infected vertices. 
Tracing and self-reporting happens with the same probability, $(1-p_s)p_t$, for all infected vertices. 
Therefore, for a given infected vertex in the tree that infects $k$ neighbors of which $d$ neighbors trace it, the first of these $d$ `tracing' contact can be viewed as the first red ball drawn without replacement from an urn with $d$ red balls and $k-d$ black balls. 
The number of black balls drawn before the first red ball is on average $(k-d)/(d+1)$, which corresponds to the average number of infectious contacts of a vertex before it is first traced. Therefore, the average fraction of non-tracing contacts that occur before the vertex is traced equals $1/(d+1)$.

\begin{figure}
    \centering
    \includegraphics[width=0.5\linewidth]{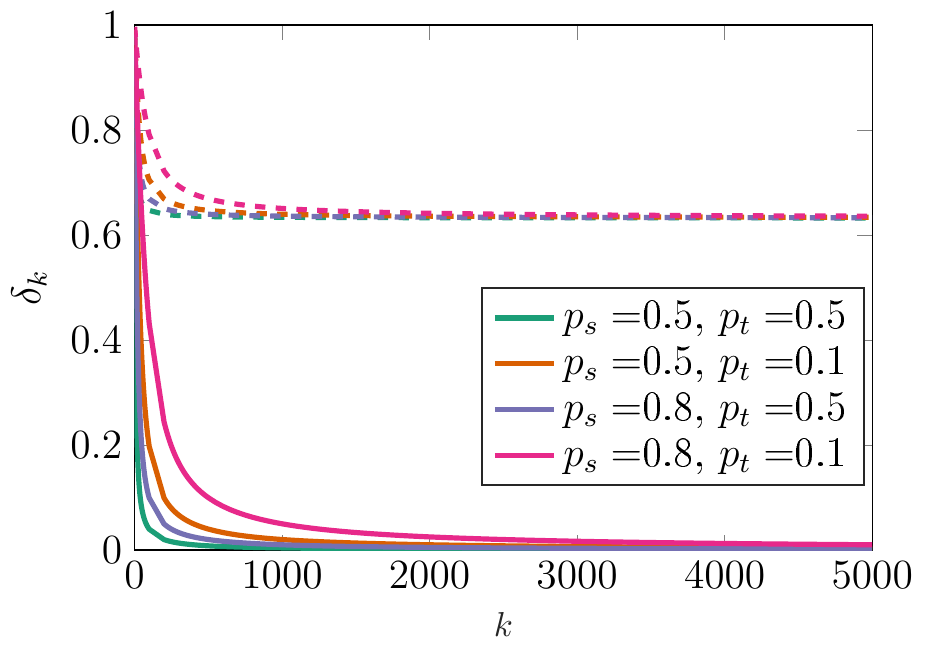}
    \caption{$\delta_k$ as a function of $k$ for various values of $p_s$ and $p_t$. The solid lines plot $\delta_k$ for immediate tracing (Eq.~\eqref{eq:deltaimmediate}), whereas the dotted lines plot $\delta_k$ for tracing with delay (Eq.~\eqref{eq:deltadelay}), using $\lambda=T=1$.}
    \label{fig:delta}
\end{figure}

The number of tracing vertices, $d$, is binomially distributed with parameters $(k, (1-p_s)p_t)$, where $k$ denotes the number of infectious contacts of the vertex. Using that $\Exp{(X+1)^{-1}}=p^{-1}(1+k)^{-1}(1-(1-p)^{k+1})$ when $X$ is distributed as Bin$(k,p)$, we obtain that the average fraction of contacts that appear before the first tracing occurs, $\delta_k$, equals 
\begin{equation}\label{eq:deltaimmediate}
\delta_k=\frac{1-\left(1- (1-p_s)p_t\right)^{k+1}}{(1-p_s)p_t(1+k)},
\end{equation}
so that $\delta_k$ is decreasing in $k$ (see Figure~\ref{fig:delta}), and asymptotically, as $k$ becomes large, we have:
\begin{equation}\label{eq:deltakasymptotic}
\delta_k = \frac{1}{ (1-p_s)p_tk}(1+o(1)). \end{equation}
Thus, we see that $\delta_k$ tends to zero when $k$ becomes large, implying that for large values of $k$, only a vanishing fraction of contacts will not be traced.

\paragraph{Phase transition under immediate tracing.}
From~\eqref{eq:deltakasymptotic} we obtain that the expected number of edges that remains for every vertex of degree $k$ is asymptotically $(1-\delta_k) k \approx 1/((1-p_s)p_t)$.
 As this quantity is independent of the vertex degree $k$, one might expect that the immediate tracing process removes the degree-heterogeneity.
 We will now show that the immediate tracing process is indeed very effective by calculating the critical value for the infectiousness parameter $\pi$, $\pi_c$ after which the epidemic outbreak becomes extensive. That is, when $\pi<\pi_c$, the size of epidemic outbreaks are sub-linear in the total number of nodes, and when $\pi>\pi_c$, this size is linear.
When the outbreak size scales linearly with the total number of vertices, we call such outbreak extensive or \emph{giant}. 

In Appendix~\ref{sec:picritdelta}, we show that there is a giant outbreak when
\begin{equation}
    \frac{g_{D}'(1-(1-p_s)p_t\pi))}{\Exp{D}}<1-\frac{(1-p_s)p_t}{p_s},
\end{equation}
where the random variable $D$ denotes the degree of a randomly chosen vertex in the network, and $g_D(x)$ its probability generating function, $g_D(x)=\sum_kq_kx^k$. 
Thus, the critical value of the percolation parameter $\pi_c$ at which a giant outbreak is such that
\begin{equation}\label{eq:pcrit}
    \frac{g_{D}'(1-(1-p_s)p_t\pi_c))}{\Exp{D}}=1-\frac{(1-p_s)p_t}{p_s}.
\end{equation}

\begin{figure}
    \centering
    \begin{subfigure}{0.45\linewidth}
    \includegraphics[width=\linewidth]{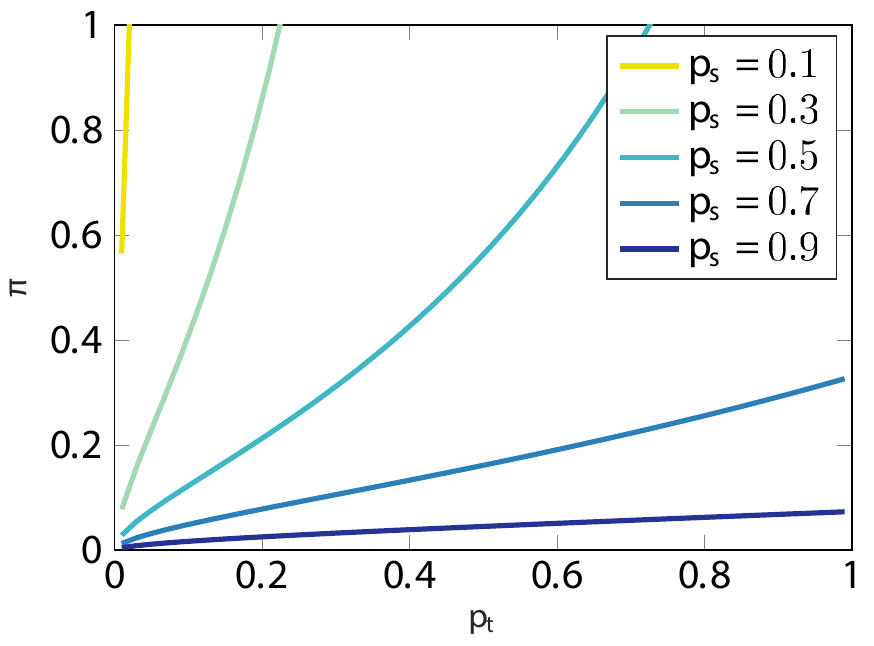}
    \caption{}
    \label{fig:critpipl}
    \end{subfigure}
    \begin{subfigure}{0.45\linewidth}
    \includegraphics[width=\linewidth]{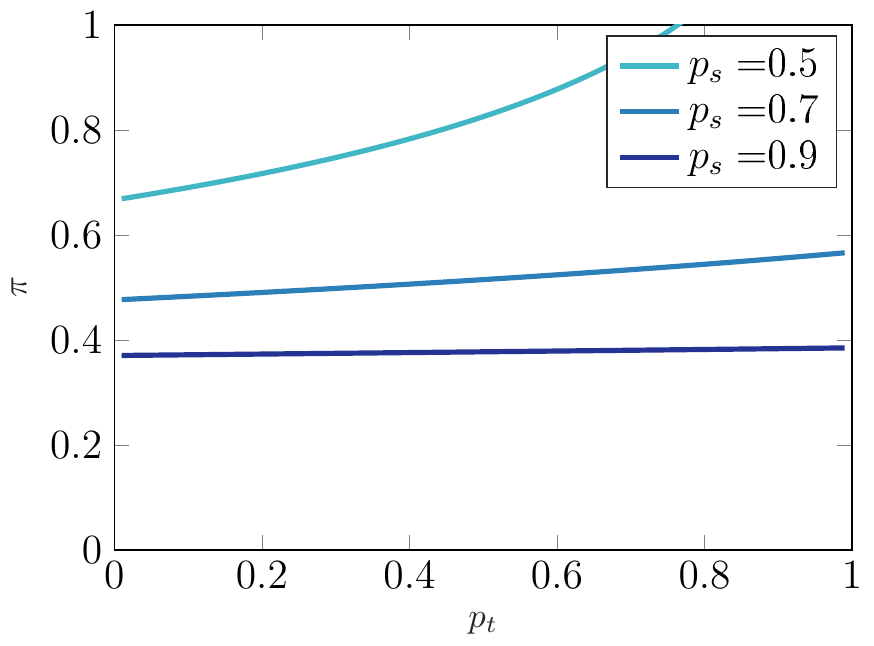}
    \caption{}
    \label{fig:critpireg}
    \end{subfigure}
    \caption{The critical percolation value $\pi_c$ from~\eqref{eq:pcrit} as a function of $p_s$ and $p_t$ in networks with (a) a power-law degree distribution with exponent 2.65 and average degree 4, (b) a regular graph of degree 4.}
    \label{fig:critpi}
\end{figure}

Figure~\ref{fig:critpi} shows the value of $\pi_c$ for two choices of the degree distribution: a regular graph where every vertex has degree 4 ($q_4=1$), and a power-law degree distribution with exponent 2.65 and average degree 4 ($q_k=Ck^{-2.65}$).
Interestingly, we see a qualitative difference between the tracing and no-tracing scenarios. 
Figure~\ref{fig:critpipl} shows that $\pi_c>0$ when $p_s,p_t>0$ even for power-law distributions with degree exponent $\tau\in(2,3)$. This means that under tracing, there is a regime for the infectiousness parameter $\pi$ such that there are only small outbreaks. On the other hand, without tracing, $\pi_c=0$ for power-law distributions with degree exponent $\tau\in(2,3)$~\cite{pastor2001}, showing that a giant outbreak always occurs regardless of the value of infectiousness $\pi$. 
Thus, this tracing process is very effective: it can reduce an extensive outbreak to have a sub-extensive size.

In the standard SIR model, a  comparable qualitative change in the size of the outbreak corresponds to a bifurcation taking place when the basic reproduction number $R_0=1$. In the regular graph, Figure~\ref{fig:critpireg}, decreasing $p_s$ or increasing $p_t$ increases the critical value $\pi_c$. Thus, when decreasing $p_s$ or increasing $p_t$, there is a wider range of values of the infectiousness parameter $\pi$ such that only small outbreaks occur, or alternatively, where the effective value of $R_0$ remains below one. 

\subsubsection{Tracing with delay}\label{sec:delay}
Even though the immediate tracing can result in a significant reduction of the giant outbreak, in practice, the tracing process may not be immediate.
 In what follows, we assume that there is a time-delay between the moment when a vertex self-reports and successfully traces its infector and the moment when the infector quarantines. We then again obtain an expression for the probability that the connection to a vertex is removed when its parent is traced, and obtain a degree-dependent version of the parameter $\delta$: $\delta_k$.

Suppose that it takes time $T$ for a vertex to self-report and trace its infector, and that all infections from a degree-$k$ vertex occur as independent exponential time clocks of rate $\lambda$. In the time-window of length $T$ (the incubation period) in which an infector is not traced yet, it can still infect others. Specifically, every remaining neighbor of the infector is infected independently in this time interval with probability $1-\textup{e}^{-\lambda T}$. 

If we denote the number of neighbors of a degree-$k$ vertex that are infected during the incubation period by $N_q$, and the number of vertices that were already infected before the incubation period started by $N_b$, then $N_q$ is distributed as a Binomial$(k-N_b,1-\textup{e}^{-\lambda T})$ variable. Thus, we obtain
\begin{align*}
    \Exp{N_q}& =\Exp{\Exp{N_q|N_b}}=(1-\textup{e}^{-\lambda T})\Exp{k-N_b}.
    \end{align*}
    We then use that $\Exp{N_b}=k\delta_k$ with $\delta_k$ as in~\eqref{eq:deltaimmediate} to obtain 
    \begin{align*}
    \Exp{N_q} = (1-\textup{e}^{-\lambda T}) k\left(1-\frac{1-(1- (1-p_s)p_t)^{k+1}}{(1-p_s)p_t(1+k)}\right)
\end{align*}
Then, the average number of vertices that are infected before tracing occurs is
\begin{align*}
    \Exp{N_q}+\Exp{N_b}& = (1-\textup{e}^{-\lambda T}) k+k\textup{e}^{-\lambda T}\frac{1-(1- (1-p_s)p_t)^{k+1}}{(1-p_s)p_t(1+k)},
\end{align*}
and the average fraction of neighbors that are infected before tracing occurs is
\begin{equation}\label{eq:deltadelay}
\delta_k(T)=1-\textup{e}^{-\lambda T} \left(1-\frac{1-(1- (1-p_s)p_t)^{k+1}}{(1-p_s)p_t(1+k)}\right). \end{equation}
For large $k$,
$$\delta_k(T)=1-\textup{e}^{-\lambda T} (1+o(1)), $$
which is independent of $k$. This implies that we can use $\delta=1-\textup{e}^{-\lambda T}$ as a proxy, instead of having a $k$-dependent $\delta$.

We therefore use a $k$-independent value of $\delta$ throughout the rest of the paper, which assumes a tracing process that is not immediate.

\begin{figure}
    \centering
    \includegraphics[width=0.5\textwidth]{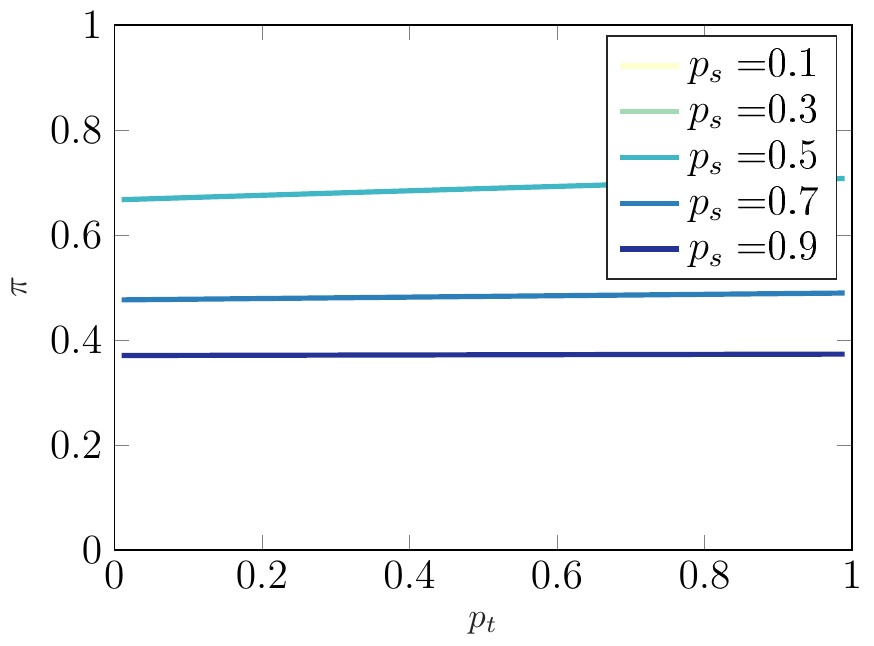}
    \caption{The critical percolation value $\pi_c$ from~\eqref{eq:picrittemp} as a function of $p_s$ and $p_t$ in networks with a regular graph degree distribution of degree 4 and $\delta=0.9$.}
    \label{fig:critpiTdependent}
\end{figure}

\paragraph{Phase transition under delayed tracing}
In Appendix~\ref{sec:picritdelayed} we show that the critical value of $\pi$ beyond which a giant outbreak occurs, satisfies 
\begin{align}\label{eq:picrittemp}
    (1-\delta)g_{D}''(1-\pi_c(1-p_s) p_t)+\delta  \Exp{D(D-1)}=\frac{\Exp{D}}{\pi_c p_s},
\end{align} 
Equation \eqref{eq:picrittemp} implies that $\pi_c=0$ for power-law degree distributions with $\tau\in(2,3)$, as then $\Exp{D^2}$, which appears on the left-hand side, diverges. 
Figure~\ref{fig:critpiTdependent} shows the value of $\pi_c$ in regular graphs. We see that the value of $\pi_c$ is more sensitive to $p_s$, the self-quarantining probability, than to $p_t$, the tracing probability. Thus, increasing the effectiveness of the tracing procedure barely influences the value of the epidemic threshold, though it may still influence the final size of the epidemic. 

The influence of the tracing process on the critical value beyond which an epidemic becomes extensive is substantially more pronounced when the tracing process is immediate. Under immediate tracing, there is a wider range of parameters where a giant outbreak becomes a sublinear outbreak (or where $R_0$ is pushed below one) than under delayed tracing. 

\section{Final outbreak size under contact tracing}\label{sec:gfanalysis}
\begin{figure}
    \centering
    \begin{subfigure}{0.45\linewidth}
    \includegraphics[width=\linewidth]{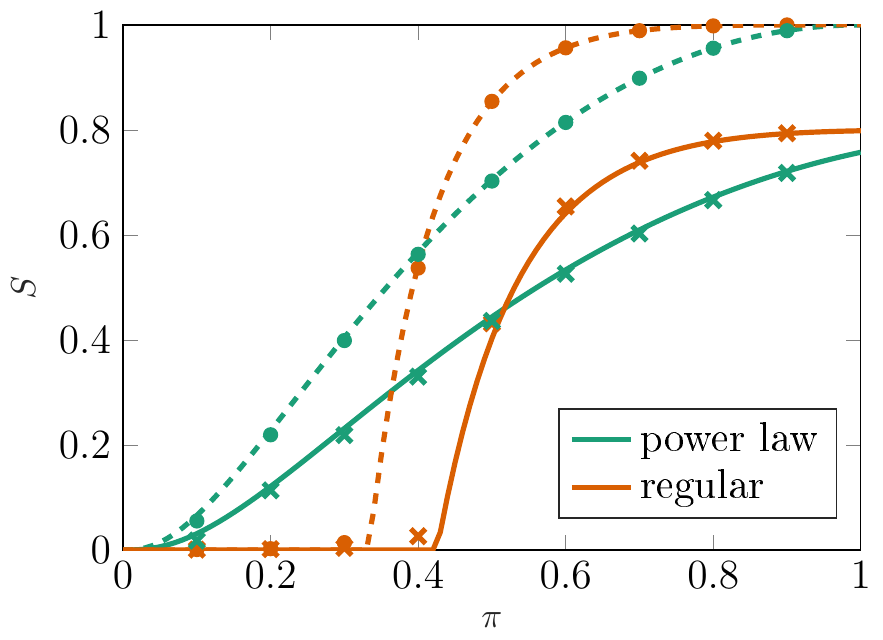}
    \caption{}
    \label{fig:S}
    \end{subfigure}
    \begin{subfigure}{0.45\linewidth}
    \includegraphics[width=\linewidth]{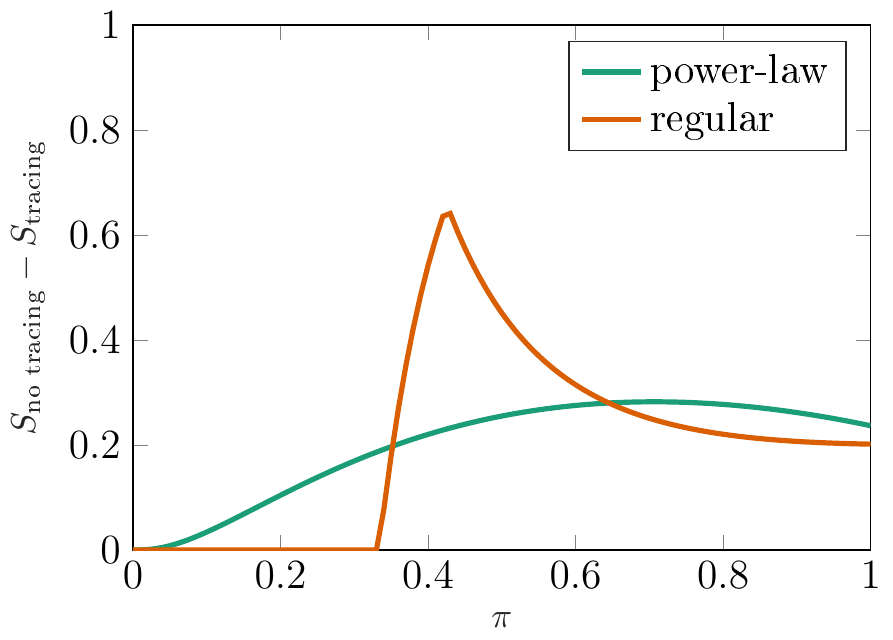}
    \caption{}
    \label{fig:effectiveness}
    \end{subfigure}
    \caption{Contact tracing with $\delta=0.9$, $p_s=0.8$, $p_t=0.6$. (a) The giant outbreak size before (dashed line) and after contact tracing (solid line) obtained from~\eqref{eq:Snotriang} in networks with a power-law degree distribution with exponent 2.65 and average degree 4, and a regular graph with degree 4. Marks are the average over 100 simulations of graphs of size $n=10.000$ (b) The effectiveness~\eqref{eq:effectiveness} of contact tracing for the power-law and the regular graph.}
    \label{fig:effheterogeneity}
\end{figure}

We now investigate the size of the remaining outbreak after tracing using a generating function approach under fixed $\delta$, as described in Section~\ref{sec:delay}. 
In Appendix~\ref{sec:computeS}, we show that in the large-network limit, the fraction of vertices in the giant outbreak $S$ is given by
\begin{equation}\label{eq:Snotriang}
S=p_s-p_sg_D(1-\pi+\pi u),
\end{equation}
where $u$ is obtained by solving the implicit equation 
\begin{align*}
u=1-p_s+p_s [&g_{D^*-1}(\delta  \pi (u-1)+1)-g_{D^*-1}\left(\left(\left(p_s-1\right) p_t+1\right) (\delta  \pi (u-1)+1)\right)\\
&+g_{D^*-1}\left((\pi (u-1)+1) \left(\left(p_s-1\right) p_t+1\right)\right)
],
\end{align*}
 where $g_{D^*-1}(x)$ is the generating function for the excess degree distribution: $g_{D^*-1}(x)=g'_D(x)/\mathbb{E}[D]$. Figure~\ref{fig:S} plots the size of the giant outbreak for networks with two different degree distributions, and shows that the analytical results of~\eqref{eq:Snotriang} match well with numerical simulations.

By comparing the outbreak size with and without tracing, we can determine the effectiveness of contact tracing. That is,
\begin{equation}\label{eq:effectiveness}
\text{eff}=S_{\text{no tracing}}-S_{\text{tracing}}, 
\end{equation}
the outbreak size in an epidemic without contact tracing, minus the outbreak size in an epidemic with tracing. Here the outbreak size without tracing can be obtained by setting $p_s=1$. Figure~\ref{fig:effectiveness} plots the effectiveness of contact tracing for two networks with the same average degree, but different degree distributions: a power-law degree distribution and a regular degree distribution. 
In both networks, the effectiveness of the tracing process depends on the infectiousness parameter $\pi$. 
In the regular network, the tracing process may shift the critical value of $\pi_c$ where the giant outbreak occurs, so that tracing completely removes a giant outbreak. In that regime, tracing is very effective. When a giant outbreak occurs in both the epidemic with tracing and in the epidemic without tracing, the effectiveness of contact tracing deceases in $\pi$. That is, the more infectious the disease, the less effective the tracing procedure. In the power-law network, a giant outbreak is always present in both the traced and the non-traced version of the epidemic. In this situation, there seems to be an `optimal' value of the infectiousness parameter $\pi$ where the tracing process is most effective. 

We see that tracing is not necessarily more effective in heterogeneous power-law networks compared to the homogeneous regular graph, in contrast with previous studies~\cite{kojaku2020,bianconi2020}. This difference is caused by the immediate tracing assumption discussed in Section~\ref{sec:delta}. Immediate tracing removes most of the degree-heterogeneity, and is therefore extremely effective on heterogeneous networks, which were studied in~\cite{kojaku2020,bianconi2020}. However, Figure~\ref{fig:effectiveness} shows that contact tracing with delay is sometimes more effective on homogeneous networks than on heterogeneous networks. For larger values of $\pi$, tracing becomes more effective on the heterogeneous power-law network than on the homogeneous regular graph.

\section{The effect of clustering on tracing}\label{sec:clustering}
 The configuration model is known to be locally tree-like: the fraction of triangles in the network vanishes asymptotically \cite{bollobas2015old}. However, many real-world networks contain a non-trivial amount of triangles, which motivates studying the tracing process on a configuration model with enhanced clustering~\cite{newman2009}. 
 In this model, each vertex $v$ has an edge-degree $d_v^{(1)}$ and a triangle degree $d_v^{(2)}$, denoting the number of triangles that the vertex is part of. Then a random graph is formed by pairing edges uniformly at random and pairing triangles uniformly at random.

Let the degree-triangle distribution be denoted by $q_{k,l}$, where $k$ denotes the edge-degree, and $l$ the triangle-degree.
 Let $g(x,y)=\sum_{k,l>0}q_{k,l}x^ky^l$ be the generating function of the edge and triangle degrees.
  Furthermore, let
\begin{align}
    g_p(x,y)& =\frac{1}{\langle k\rangle}\sum_{k,l>0}sq_{k,l}x^{k-1}y^l,\\
    g_q(x,y)& =\frac{1}{\langle l\rangle}\sum_{k,l>0}tq_{k,l}x^ky^{l-1},
\end{align}
with
$$
\langle s\rangle := \sum_{k,l>0}kq_{k,l},\quad\langle l\rangle:=\sum_{k,l>0}lq_{k,l},
$$
be the generating functions of the number of edges and triangles that are reached by following a randomly chosen edge and a randomly chosen triangle respectively.

In Appendix~\ref{sec:clusterderiv}, we show that the outbreak size after tracing equals 
\begin{equation}\label{eq:Striang}
    S=p_s-p_sg(1-\pi+\pi u,(1-\pi)^2+2(1-\pi)^2\pi v +\pi^2(3-2\pi)v^2),
\end{equation}
where $u$ and $v$ are obtained by solving the system of  implicit equations
\begin{align*}
    u&=1-p_s+p_s\Bigg(g_p(1-\pi+\pi w u,(1-\pi)^2+2(1-\pi)^2\pi w v + 2\pi^2(1-\pi)v^2w +\pi^2w^2v^2)\\
    & \quad + g_p\left(1-\pi\delta+\pi\delta u,(1-\pi\delta)^2+2(1-\pi\delta)^2\pi\delta v +\pi^2\delta^2(3-2\pi\delta)v^2\right)\\
    & \quad - g_p\bigg(1-\pi+\pi w(1-\delta+\delta u),(\pi +\pi  (\delta -1) w-1)^2+2 \pi  \delta  w (\pi  \delta -1) (\pi +\pi  (\delta -1) w-1)v\\
    & \quad\quad -w\pi ^2 \delta ^2(2 (\pi -1)+w (2 \pi  (\delta -1)-1))v^2\bigg)\Bigg)
\end{align*}
and
\begin{align*}
    v&=1-p_s+p_s\Bigg(g_q(1-\pi+\pi w u,(1-\pi)^2+2(1-\pi)^2\pi w v +\pi^2(1-\pi)w^2v^2)\\
    & \quad + g_q\left(1-\pi\delta+\pi\delta u,(1-\pi\delta)^2+2(1-\pi\delta)^2\pi\delta v +\pi^2\delta^2(3-2\pi\delta)v^2\right)\\
    & \quad - g_q\bigg(1-\pi+\pi w(1-\delta+\delta u),(\pi +\pi  (\delta -1) w-1)^2 +2 \pi  \delta  w (\pi  \delta -1) (\pi +\pi  (\delta -1) w-1)v\\
    & \quad\quad -w\pi ^2 \delta ^2 (2 (\pi -1)+w (2 \pi  (\delta -1)-1))v^2\bigg)\Bigg),
\end{align*}
where $w=p_s+(1-p_s)(1-p_t)$.

\begin{figure}
\centering
\begin{subfigure}[b]{0.48\textwidth}
    \centering
    \includegraphics[width=\textwidth]{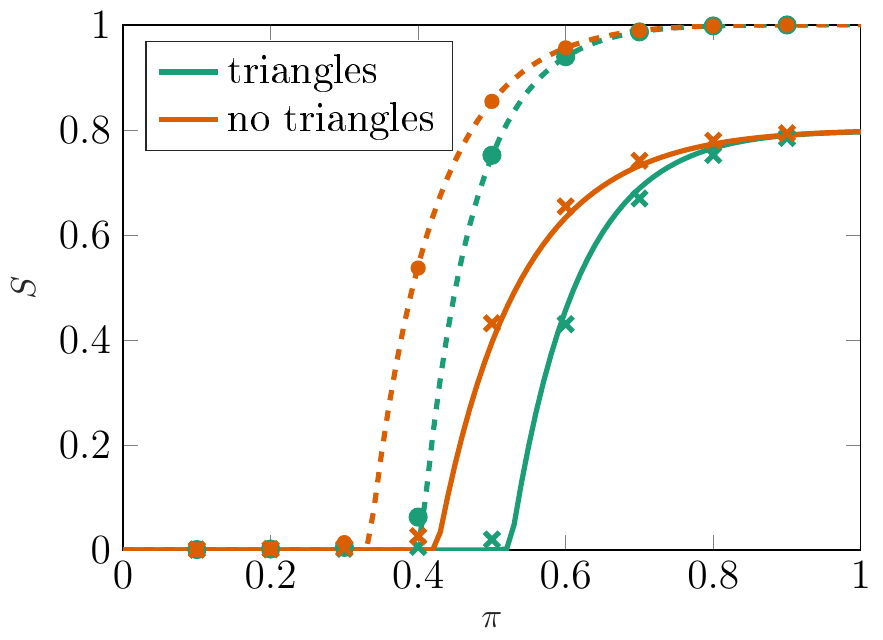}
    \caption{}
    \label{fig:Striangreg}
\end{subfigure}
\begin{subfigure}[b]{0.48\textwidth}
    \centering
    \includegraphics[width=\textwidth]{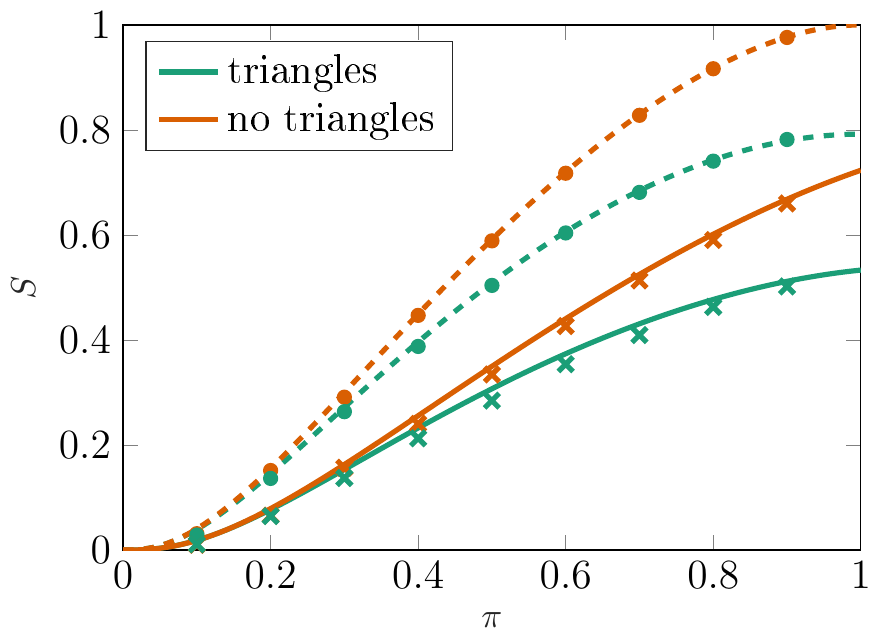}
    \caption{}
    \label{fig:Striangpl}
    \end{subfigure}
    \caption{The giant outbreak size before (dashed line) and after contact tracing (solid line) obtained from~\eqref{eq:Striang} with $\delta=0.9$, $p_s=0.8$, $p_t=0.6$ in (a) a regular graph with edge-degree 4, and triangle degree 0 (orange) and a regular graph with triangle-degree 2 and edge-degree 0 (green) and in (b) a graph with power-law edge-degrees with exponent 2.65, average edge-degree 3.4, and triangle degree 0 (orange) and a graph with power-law triangle-degree with exponent 2.65, average 1.7 and edge-degree 0. Marks are averages over 100 simulations of networks with $n=10.000$. }
\end{figure}

Figures~\ref{fig:Striangreg} and~\ref{fig:Striangpl} show the epidemic size in networks with the same degree distribution but with a different amount of triangles. The analytic results for the final epidemic outbreak on networks with triangles obtained from~\eqref{eq:Striang} closely matches the results obtained by numerical simulations.

\begin{figure}
\centering
\begin{subfigure}[b]{0.48\textwidth}
    \centering
    \includegraphics[width=\textwidth]{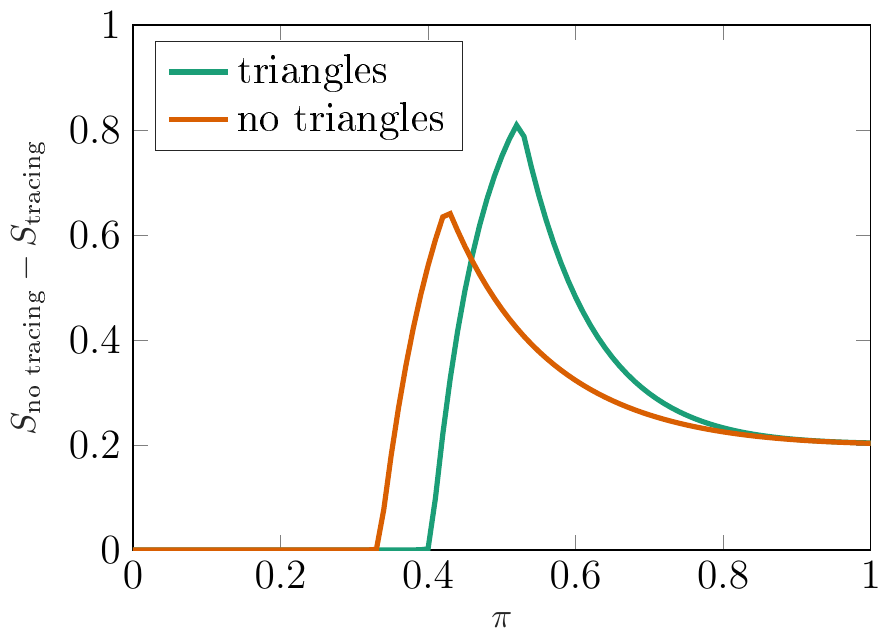}
    \caption{}
    \label{fig:efftriangreg}
\end{subfigure}
\begin{subfigure}[b]{0.48\textwidth}
    \centering
    \includegraphics[width=\textwidth]{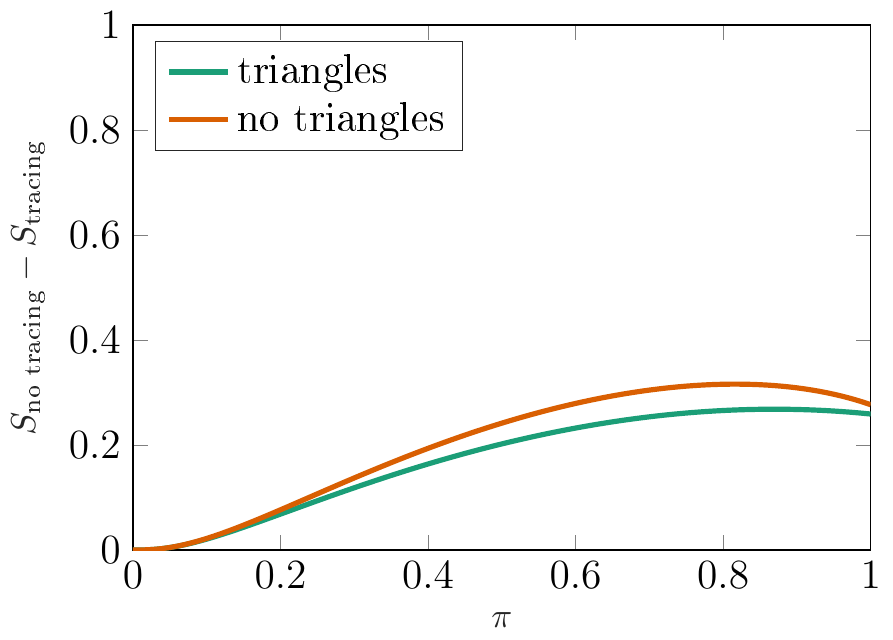}
    \caption{}
    \label{fig:efftriangpl}
    \end{subfigure}
    \caption{The effectiveness of contact tracing with $\delta=0.9$, $p_s=0.8$, $p_t=0.6$ in (a) the networks of Figure~\ref{fig:Striangreg} and (b) the networks of Figure~\ref{fig:Striangpl}. }
    \label{fig:efftriang}
\end{figure}

Furthermore, one may conclude from Figures~\ref{fig:efftriangreg} and~\ref{fig:efftriangpl} that the effectiveness of the contact tracing non-trivially depends on the amount of clustering. In the regular graph, Figure~\ref{fig:efftriangreg} shows that there is a range of the infectiousness parameter $\pi$ where the tracing procedure is more effective on clustered networks than on tree-like networks, but there is also a range of parameters where the tracing procedure is more effective on the tree-like networks instead. On the heterogeneous power-law networks on the other hand, Figure~\ref{fig:efftriangpl} shows that the effectiveness of tracing is always higher in the tree-like network than in the clustered networks. 
Furthermore, the difference between the clustered and non-clustered networks is less pronounced in the power-law network. 

Intuitively, introducing triangles has two effects: on the one hand they make it easier for an epidemic to spread, as they induce multiple paths for a person $i$ to infect another person $j$, but on the other hand, they reduce the number of vertices that the epidemic can reach from a given vertex in $k$ steps compared to a tree. The latter effect makes it easier for the tracing process to stop the epidemic in the presence of triangles. For power-law vertices, this is less pronounced, as in the presence of high-degree vertices, it is likely that the vertex has already infected many other neighbors before being traced. This may intuitively explain the difference between introducing triangles in power-law networks compared to homogeneous networks.

In general, Figure~\ref{fig:efftriang} shows that the effectiveness of contact tracing delicately depends on the interplay between the network degree distribution and its structure in terms of clustering.


\section{Conclusion}
In this paper, we have analytically studied a contact tracing process on networks with arbitrary degree distributions. In this process, infected vertices self-report and quarantine with some probability $1-p_s$, and they trace their parent with probability $p_t$. Using generating functions, we derive analytical expressions of the giant outbreak size after the tracing process. 

We investigated the effect of the network structure on the tracing process and found that degree heterogeneity may either enhance or diminish the effectiveness of tracing depending on the exact parameter values. 
In our tracing model, we assume that there is a time-delay between the time that a person is infected and the time that its infector is traced. This assumption makes the network heterogeneity non-trivially affect the tracing effectiveness. 

Likewise, enhancing clustering in the network has a non-trivial effect on the effectiveness of contact tracing. Depending on the infectiousness of the epidemic, clustering may either increase or decrease the effectiveness of contact tracing, in contrast with conclusions from simulations on homogeneous networks~\cite{kiss2005}. This underlines the importance of taking the network structure into account when investigating such tracing processes.

In this paper, we investigated bond percolation, which can be mapped to the final size of an SIR epidemic with constant recovery duration. In further research, it would be interesting to investigate the entire time evolution of the number of infected vertices in the SIR process as well, and investigate the effect of the network structure on this time evolution. This would enable to answer the question whether the network structure affects the speed at which tracing processes slow down the spread of an epidemic. 

Furthermore, our results on power-law networks suggest that there is an optimal value of the infectiousness parameter $\pi$ such that tracing is the most effective. It would be interesting to investigate the relation between this optimal value of $\pi$ and the parameters of the tracing process, to enable the design of optimally efficient tracing processes.

\bibliographystyle{abbrv}
\bibliography{references}

\appendix
\section{Derivation of the critical value for immediate tracing}\label{sec:picritdelta}
We model the infection tree as a branching process with a certain offspring distribution. A giant component emerges when the average offspring surpasses one. Thus, we calculate the average number of non-self-reporting offspring of a vertex $N_f$ (as the self-reporting vertices will have zero offspring, and therefore do not contribute to creating a giant outbreak).



Let $N_t$ the total number of offspring of a vertex in the second tier of the branching process in the original network. Then, $N_t$ is distributed as $D^*-1$, where $D^*$ denotes the size-biased degree distribution. Let $N_t^{(\pi)}$ be the number of neighbors after percolation with parameter $\pi$, so that $N_t^{(\pi)}=\text{Bin}(N_t,\pi)$. Let $R$ denote the number of `reporting' neighbors of the vertex, so that $R=\text{Bin}(N_t^{\pi},1-p_s)$. Finally, let $R_t$ the number of reporting vertices that also 'trace' its parent vertex, so that $R_t=\text{Bin}(R,p_t)$. Given $R_t, R, N_t^{(\pi)}$, the vertex infects $N_t^{(\pi)}-R$ non-reporting vertices before taking tracing into account. On average, $(N_t^{(\pi)}-R)/(R_t+1)$ of these will remain after tracing (so when permuting randomly, and removing every neighbor after the first `tracing' sibling. Thus,
\begin{align*} 
\Exp{N_f} & =\Exp{\Exp{N_f\mid N_t^{(\pi)},R}}=\Exp{\Exp{\Exp{N_f\mid R_t}\mid  N_t^{(\pi)}}}\\
& =\Exp{\Exp{\frac{N_t^{(\pi)}-R}{R_r+1}\mid  N_t^{(\pi)},R}}
\end{align*}
Using that $\Exp{(X+1)^{-1}}=p^{-1}(1+k)^{-1}(1-(1-p)^{k+1})$ when $X$ is distributed as Bin$(k,p)$, we obtain
\begin{align*}
 \Exp{N_f} &  = \Exp{\Exp{\frac{(N_t^{(\pi)}-R)(1-(1-p_t)^{R+1})}{(R+1)p_t}\mid  N_t^{(\pi)}}}\\
& = \Exp{\frac{p_s(1-(1-(1-p_s)p_t)^{N_t^{(\pi)}})}{(1-p_s)p_t}}\\
& = \Exp{\Exp{\frac{p_s(1-(1-(1-p_s)p_t)^{N_t^{(\pi)}})}{(1-p_s)p_t}\mid N_t}}
\end{align*}
Further, using that the probability generating function of a Bin$(k,p)$ random variable is $(1-p+px)^k$, we obtain
\begin{align*}
\Exp{N_f} & = \Exp{\frac{p_s(1-(1-(1-p_s)p_t\pi)^{N_t})}{(1-p_s)p_t}}\\
& = \frac{p_s}{(1-p_s)p_t}-\frac{p_s}{(1-p_s)p_t}\Exp{(1-(1-p_s)p_t\pi)^{N_t}}\\
&=\frac{p_s(1-g_{D^*-1}(1-(1-p_s)p_t\pi))}{(1-p_s)p_t},
\end{align*}
where $g_{D^*-1}(x)$ is the probability generating function of the size-biased degree distribution minus 1, so that $g_{D^*-1}(x)=g'(x)/\Exp{D}$.

A giant outbreak occurs when the expected number of offspring surpasses one, so when
\begin{equation}
    \frac{g_{D}'(1-(1-p_s)p_t\pi))}{\Exp{D}}<1-\frac{(1-p_s)p_t}{p_s}.
\end{equation}
Thus, the critical value of the percolation parameter $\pi_c$ is such that
\begin{equation}
    \frac{g_{D}'(1-(1-p_s)p_t\pi_c))}{\Exp{D}}=1-\frac{(1-p_s)p_t}{p_s}.
\end{equation}

\section{Critical value under delayed tracing}\label{sec:picritdelayed}
We now derive the critical percolation value under delayed tracing with fixed $\delta$. We use the same notation as in Appendix~\ref{sec:picritdelta}.
When $R$ vertices report themselves, the probability that their infector is not traced is $(1-p_t)^R$. There are $N_t^{(\pi)}-R$ non-reporting vertices. When their infector is traced, on average a fraction of $\delta$ of them remain infected.  Therefore,
\begin{align*} 
\Exp{N_f} & =\Exp{\Exp{N_f\mid N_t^{(\pi)},R}}\\
& =\Exp{\Exp{(1-p_t)^R(N_t^{(\pi)}-R)+(1-(1-p_t)^R)(N_t^{(\pi)}-R)\delta\mid N_t^{(\pi)}}}\\
& = \Exp{N_t^{(\pi)}p_s(1-(1-p_s) p_t)^{N_t^{(\pi)}-1}}+\delta \pi p_s\Exp{D^*-1}\\
& \quad -\delta \Exp{N_t^{(\pi)}p_s(1-(1-p_s) p_t)^{N_t^{(\pi)}-1}},
\end{align*}
where the last step used that given $N_t^{(\pi)}$, $N_t^{(\pi)}-R$ is binomially distributed with parameters $N_t^{(\pi)}$ and $(1-p_s)$, and the probability generating function of a Bin$(k,p)$ random variable is given by $(1-p+px)^k$. Also, given $N_t$, $N_t^{(\pi)}$ is binomially distributed with parameters $N_t$ and $\pi$. Thus,
\begin{align*}
\Exp{N_f} & = p_s(1-\delta)\Exp{\pi N_t(1-\pi(1-p_s) p_t)^{N_t-1}}+\delta \pi p_s\Exp{D^*-1}\\
& =p_s\pi (1-\delta)g_{D^*-1}'(1-\pi(1-p_s) p_t)+\delta \pi p_s\Exp{D^*-1}.
\end{align*}
Because $\mathbb{E}[Yx^{Y-1}]=g'_Y(x)$ for any random variable $Y$, and $N_t$ is distributed as $D^*-1$, where $D^*$ is the size-biased degree-distribution,
Finally, using that $g_{D^*-1}(x)=g'_D(x)/\mathbb{E}[D]$ and that $\mathbb{E}[D^*-1]=\mathbb{E}[D(D-1)]$, we obtain
\begin{align*}
\Exp{N_f} & = \frac{p_s\pi (1-\delta)g_{D}''(1-\pi(1-p_s) p_t)}{\Exp{D}}+\delta \pi p_s\Exp{D(D-1)}.
\end{align*}
The critical value of $\pi$ where a giant outbreak occurs, is when $\mathbb{E}[N_f]=1$, which yields equation~\eqref{eq:picrittemp}.

\section{The giant outbreak size}\label{sec:computeS}
In this section, we compute the giant outbreak size after tracing.
A vertex does not trace its infector if it does not self-report, which happens with probability $p_s$, or if it does self-report, but does not successfully trace its infector, which happens with probability $(1-p_s)(1-p_t)$.
Thus, the probability that a vertex of degree $k$ is traced by none of its offspring equals
\begin{equation}
    \Prob{\text{degree $k$ vertex not traced}}=(p_s+(1-p_s)(1-p_t))^k.
\end{equation}
Let $$p_k = \frac{(k+1)q_{k+1}}{\sum_{k\geq1}{kq_{k}}}$$
 be the excess degree distribution, and $p_k^*$ be the excess degree distribution after tracing. As a vertex loses all its offspring after self-reporting which happens with probability $1-p_s$, $p(0)$ is given by
$$p(0)=1-p_s.$$
When a vertex is not traced, its degree remains the same. When a vertex is traced, an extra layer of percolation occurs with parameter $\delta$. Thus,
\begin{align*}
p_k^*& = \underbrace{p_s}_\text{Not quarantined} \Big (  \sum_{j=k}^\infty p_{k,j}^\delta \underbrace{( 1- ( p_s + (1-p_s) (1-p_t) )^j)}_\text{At least one offspring traces this node}  \\
& \quad +  \underbrace{( p_s + (1-p_s) (1-p_t) )^k}_\text{None of the offsprings trace this node} p_k  \Big),k>0
\end{align*}
where $p_{k,j}^\delta$ is the probability that a vertex of degree $j$ has remaining degree $k$ after percolation with bond occupancy $\delta$. The generating function for $p_k^*$ is then given by:
\begin{align*}
\sum_{k=0}^\infty p_k^*x^k& =1-p_s+p_s\sum_{k=1}^\infty\sum_{j=k}^\infty p_{k,j}^\delta x^k- p_s\sum_{k=1}^\infty \sum_{j=k}^\infty x^k( p_s + (1-p_s) (1-p_t) )^j)p_{k,j}^\delta\\
&\quad +p_s\sum_{k=1}^\infty(x( p_s + (1-p_s) (1-p_t) ))^kp_k.
\end{align*}
Now
\begin{equation*}
    \sum_{k=1}^\infty(x( p_s + (1-p_s) (1-p_t) ))^kp_k=g_{D^*-1}(x( p_s + (1-p_s) (1-p_t) )),
\end{equation*}
where $g_{D^*-1}(x)$ denotes the generating function of $p_k$. Furthermore,
\begin{align*}
    \sum_{k=1}^\infty \sum_{j=k}^\infty p_{k,j}^\delta x^k& = \sum_{k=1}^\infty \sum_{j=k}^\infty x^k p_j\Prob{\text{Bin}(j,\delta)=k}= \sum_{j=1}^\infty p_j\sum_{k=0}^j x^k\Prob{\text{Bin}(j,\delta)=k}\\
    & = \sum_{j=1}^\infty p_j(1-\delta+\delta x)^j = g_{D^*-1}(1-\delta+\delta x).
\end{align*}
Similarly,
\begin{align*}
   \sum_{k=1}^\infty \sum_{j=k}^\infty x^k( p_s + (1-p_s) (1-p_t) )^j)p_{k,j}^\delta =  g_{D^*-1}\left((1-\delta+\delta x)(p_s + (1-p_s) (1-p_t))\right).
\end{align*}
Thus, 
\begin{align}\label{eq:gentracing}
\sum_{k=0}^\infty p_k^*x^k& =
1-p_s+p_s \big[
g_{D^*-1}(1 - \delta + \delta x)
-g_{D^*-1}\left((1 - \delta + \delta x) \left(\left(1-p_s\right) \left(1-p_t\right)+p_s\right)\right)\nonumber\\
& \quad +g_{D^*-1}\left(x \left(\left(1-p_s\right) \left(1-p_t\right)+p_s\right)\right)
\big].
\end{align}
This is the generating function of the degree distribution of a tracing process on a network with excess degree distribution $p_k$. However, before the tracing process takes place, an epidemic modeled by a bond percolation process with occupancy $\pi$ takes place. Thus, to obtain the generating function $G(x)$ of the degree distribution after the epidemic and the tracing process, we add the bond percolation process with bond occupancy probability $\pi$ by substituting  $x\to1 - \pi + \pi x$ in~\eqref{eq:gentracing}:
\begin{align*}
G(x)=1-p_s+p_s [&g_{D^*-1}(\delta  \pi (x-1)+1)\\
&-g_{D^*-1}\left(\left(\left(p_s-1\right) p_t+1\right) (\delta  \pi (x-1)+1)\right)\\
&+g_{D^*-1}\left((\pi (x-1)+1) \left(\left(p_s-1\right) p_t+1\right)\right).
]
\end{align*}
We then obtain the size of the giant outbreak $S=p_s-p_sg_D(1-\pi+\pi u)$, where $u$ is obtained by solving the implicit equation $u=G(u)$ and $g_D(x)$ is the generating function of the degree distribution.

By studying fixed points of the map, $s(x)=x G(s(x))$, we find the following percolation condition:
$$ \pi p_s (\delta -1) \left(\left(p_s-1\right) p_t+1\right) G'\left(\left(p_s-1\right) p_t+1\right)-\delta \pi p_s G'(1)+1=0$$

\section{Derivation of the giant outbreak size in clustered networks}\label{sec:clusterderiv}
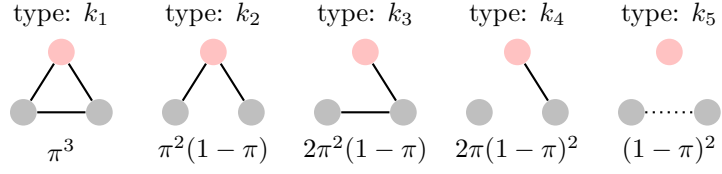
\begin{figure}
    \centering
    \begin{tikzpicture}
    \tikzstyle{vertexbas}=[circle,fill=black!25,minimum size=10pt,inner sep=0pt]
\tikzstyle{vertex} = [vertexbas, fill=red!24]
\tikzstyle{edge} = [draw,thick,-]
\node[vertexbas] (a) at (0,0) {};
\node[vertexbas] (b) at (1,0) {};
\node[vertex] (c) at (0.5,0.8) {};
\node at (0.5,-0.5) {$\pi^3$};
\node at (0.5,1.3) {type: $k_1$};
\path[edge] (a)--(b)--(c)--(a);
\node[vertexbas] (a2) at (2,0){};
\node[vertexbas] (b2) at (3,0) {};
\node[vertex] (c2) at (2.5,0.8) {};
\path[edge] (b2)--(c2)--(a2);
\node at (2.5,-0.5) {$\pi^2(1-\pi)$};
\node at (2.5,1.3) {type: $k_2$};
\node[vertexbas] (a3) at (4,0){};
\node[vertexbas] (b3) at (5,0) {};
\node[vertex] (c3) at (4.5,0.8) {};
\path[edge] (a3)--(b3)--(c3);
\node at (4.5,-0.5) {$2\pi^2(1-\pi)$};
\node at (4.5,1.3) {type: $k_3$};
\node[vertexbas] (a4) at (6,0){};
\node[vertexbas] (b4) at (7,0) {};
\node[vertex] (c4) at (6.5,0.8) {};
\path[edge] (b4)--(c4);
\node at (6.5,-0.5) {$2\pi(1-\pi)^2$};
\node at (6.5,1.3) {type: $k_4$};
\node[vertexbas] (a5) at (8,0){};
\node[vertexbas] (b5) at (9,0) {};
\node[vertex] (c5) at (8.5,0.8) {};
\path[edge,dotted] (a5)--(b5);
\node at (8.5,-0.5) {$(1-\pi)^2$};
\node at (8.5,1.3) {type: $k_5$};
    \end{tikzpicture}
    \caption{After percolation with probability $\pi$, a triangle that is reached at the red vertex has become one of these types. Thus, when arriving at a percolated triangle at the red vertex, zero, one, or two other vertices may be reached. The labels below the types provide the probability that a percolated triangle equals this type.}
    \label{fig:triangperc}
\end{figure}

Under bond percolation with probability $\pi$, a triangle from a given vertex can still be connected to its two triangle members, with probability $\pi^2(3-2\pi)$, it can connect to only one of its triangle members, with probability $2(1-\pi)^2\pi$, or it can become disconnected from both other triangle members, with probability $(1-\pi)^2$. 
Thus, for a vertex of triangle-degree $k$, the number of neighbors that are reachable through these triangles after bond percolation, has generating function $g_{D^*-1}(z)=((1-\pi)^2+2(1-\pi)^2\pi z+\pi^2(3-2\pi)z^2)^k$. Let $u$ denote the probability that a randomly chosen half-edge is not connected to the giant component. Similarly, let $v$ denote the probability that following a randomly chosen triangle does not lead to the largest component.
Then, after bond percolation with probability $\pi$, 
\begin{align}
    u& =g_p(1-\pi+\pi u,(1-\pi)^2+2(1-\pi)^2\pi v +\pi^2(3-2\pi)v^2),\\
    v& =g_q(1-\pi+\pi u,(1-\pi)^2+2(1-\pi)^2\pi v +\pi^2(3-2\pi)v^2).
\end{align}
Adding site percolation with probability $p_s$ results in
\begin{align}
    u& =1-p_s+p_sg_p(1-\pi+\pi u,(1-\pi)^2+2(1-\pi)^2\pi v +\pi^2(3-2\pi)v^2),\\
    v& =1-p_s+p_sg_q(1-\pi+\pi u,(1-\pi)^2+2(1-\pi)^2\pi v +\pi^2(3-2\pi)v^2).
\end{align}
Let $w$ denote that a vertex of degree $1$ is traced by none of its offspring, so that
\begin{equation}
    \Prob{\text{degree $k$ vertex not traced}}=(p_s+(1-p_s)(1-p_t))^k=w^k.
\end{equation}
When a vertex is traced, an extra layer of percolation with parameter $\delta$ takes place, so that combined, this is percolation with parameter $\pi\delta$. However, the probability of this taking place, depends on the degree of the vertex after the first layer of percolation with parameter $\pi$. After the first layer of percolation with parameter $\pi$, triangles are percolated into 5 possible types, as illustrated in Figure~\ref{fig:triangperc}. In the leftmost two types, the percolated triangle contributes with two to the degree of the red vertex, the rightmost percolated triangle adds zero to the degree of the red vertex, and in the other two types, the percolated triangle adds one to the degree of the vertex. Thus, when we denote the number of percolated triangles of these types by $k_1,k_2,\dots,k_5$, see Figure~\ref{fig:triangperc}, the degree of the vertex from the percolated triangles equals $2(k_1+k_2)+k_3+k_4$. The number of vertices that are reached through these percolated triangles equals $2(k_1+k_2+k_3)+k_4$.

Let $\hat{D}^{(1)}$ and $\hat{D}^{(2)}$ denote the degree of the number of edges and triangle-edges respectively after the tracing process. Furthermore, let $k_6$ denote the number of remaining half-edges attached to a vertex after percolating the half-edges with probability $\pi$.
As the probability of a vertex not being traced equals $w$ to the power of the degree after percolation with parameter $\pi$, 
\begin{align*}
    \Exp{x^{\hat{D}^{(1)}}y^{\hat{D}^{(2)}}\mathbbm{1}_{\text{not traced}}}&=\Exp{\Exp{x^{\hat{D}^{(1)}}y^{\hat{D}^{(2)}}\mathbbm{1}_{\text{not traced}}\mid k_1,\dots,k_5,k_6}}\\
    &=\Exp{x^{k_6}y^{2(k_1+k_2+k_3)+k_4}w^{2(k_1+k_2)+k_3+k_4+k_6}}\\
    & =g_p(1-\pi+\pi w x,(1-\pi)^2+2(1-\pi)^2\pi w y \\
    & \quad + 2\pi^2(1-\pi)y^2w +\pi^2w^2y^2),
\end{align*}
where the last step used the probabilities in Figure~\ref{fig:triangperc}, and the generating function of the multinomial distribution. Also, when a vertex is traced, its neighbors are percolated with parameter $\delta$. Therefore,
\begin{align*}
    & \Exp{x^{\hat{D}^{(1)}}y^{\hat{D}^{(2)}}\mathbbm{1}_{\text{traced}}}=\Exp{\Exp{x^{\hat{D}^{(1)}}y^{\hat{D}^{(2)}}\mathbbm{1}_{\text{traced}}\mid k_1,\dots,k_5,k_6}}\\
    &=\Exp{(1-w^{2(k_1+k_2)+k_3+k_4+k_6})\Exp{x^{\hat{D}^{(1)}}y^{\hat{D}^{(2)}}\mid k_1,\dots,k_5,k_6, \text{traced}}}\\
    & =\Exp{x^{{D}^{(1,\pi\delta)}}y^{{D}^{(2,\pi\delta)}}}-\Exp{w^{2(k_1+k_2)+k_3+k_4+k_6}\Exp{x^{\hat{D}^{(1)}}y^{\hat{D}^{(2)}}\mid k_1,\dots,k_5,k_6, \text{traced}}},
\end{align*}
where $D^{(1,\pi\delta)}$ and $D^{(2,\pi\delta)}$ denote the degree and triangle-degree respectively of a vertex after percolation with parameter $\pi\delta$. Thus,
\begin{align*}
    \Exp{x^{{D}^{(1,\pi\delta)}}y^{{D}^{(2,\pi\delta)}}}= & g_p(1-\pi\delta+\pi\delta x,(1-\pi\delta)^2+2(1-\pi\delta)^2\pi\delta y +\pi^2\delta^2(3-2\pi\delta)y^2).
\end{align*}
For the second term, we have to take into account that the percolated triangles of Figure~\ref{fig:triangperc} are again percolated with parameter $\delta$. Let $\phi_{t,i}$ denote the probability that a percolated triangle of type $k_t$ reaches $i$ neighbors after an extra layer of percolation with probability $\delta$. For example, $\phi_{1,0}=(1-\delta)^2$, the probability that a full triangle does not reach both its neighbors after percolation with parameter $\delta$. Furthermore, let $\zeta_t$ denote the probability that after percolation of a full triangle with probability $\pi$, the percolated triangle is of type $k_t$. The $\zeta_t$ are given in Figure~\ref{fig:triangperc}, and for example $\zeta_2=\pi^2(1-\pi)$. Then we obtain
\begin{align*}
   &\Exp{w^{2(k_1+k_2)+k_3+k_4+k_6}\Exp{x^{\hat{D}^{(1)}}y^{\hat{D}^{(2)}}\mid k_1,\dots,k_5,k_6, \text{traced}}}\\
   & = \Exp{w^{2(k_1+k_2)+k_3+k_4+k_6}(1-\delta+\delta x)^{k_6}\prod_{t=1}^5(\sum_{i=0}^2\phi_{t,i}y^i)^{k_t}}\\
   & = g_p\left(1-\pi+\pi w(1-\delta+\delta x),\sum_{t=1}^5\zeta_t w^{a_t}\sum_{i=0}^2\phi_{t,i}y^i\right),
\end{align*}
where $a_t=2$ for $t=1,2$, $a_t=1$ for $t=3,4$ and $a_t=0$ for $w=5$. Plugging in the expressions for $\zeta_t$ and $\phi_{t,i}$ and simplifying, yields
\begin{align*}
   &\Exp{w^{2(k_1+k_2)+k_3+k_4+k_6}\Exp{x^{\hat{D}^{(1)}}y^{\hat{D}^{(2)}}\mid k_1,\dots,k_5,k_6, \text{traced}}}\\
   &= g_p\bigg(1-\pi+\pi w(1-\delta+\delta u),(\pi +\pi  (\delta -1) w-1)^2 +2 \pi  \delta  w (\pi  \delta -1) (\pi +\pi  (\delta -1) w-1)v\\
    & \quad\quad -w\pi ^2 \delta ^2(2 (\pi -1)+w (2 \pi  (\delta -1)-1))v^2\bigg)\Bigg)
\end{align*}
Thus, when we let $u$ denote the probability that a vertex that is reached by following a randomly chosen half-edges is not connected to the giant component, we obtain
\begin{align*}
    u&=1-p_s+p_s\Bigg(g_p(1-\pi+\pi w u,(1-\pi)^2+2(1-\pi)^2\pi w v + 2\pi^2(1-\pi)v^2w +\pi^2w^2v^2)\\
    & \quad + g_p(1-\pi\delta+\pi\delta u,(1-\pi\delta)^2+2(1-\pi\delta)^2\pi\delta v +\pi^2\delta^2(3-2\pi\delta)v^2)\\
    & \quad - g_p\bigg(1-\pi+\pi w(1-\delta+\delta u),(\pi +\pi  (\delta -1) w-1)^2+2 \pi  \delta  w (\pi  \delta -1) (\pi +\pi  (\delta -1) w-1)v\\
    & \quad\quad -w\pi ^2 \delta ^2(2 (\pi -1)+w (2 \pi  (\delta -1)-1))v^2\bigg)\Bigg)
\end{align*}
Similarly 
\begin{align*}
    v&=1-p_s+p_s\Bigg(g_q(1-\pi+\pi w u,(1-\pi)^2+2(1-\pi)^2\pi w v +\pi^2(1-\pi)w^2v^2)\\
    & \quad + g_q(1-\pi\delta+\pi\delta u,(1-\pi\delta)^2+2(1-\pi\delta)^2\pi\delta v +\pi^2\delta^2(3-2\pi\delta)v^2)\\
    & \quad \quad- g_q\bigg(1-\pi+\pi w(1-\delta+\delta u),(\pi +\pi  (\delta -1) w-1)^2 +2 \pi  \delta  w (\pi  \delta -1) (\pi +\pi  (\delta -1) w-1)v\\
    & \quad\quad -w\pi ^2 \delta ^2 (2 (\pi -1)+w (2 \pi  (\delta -1)-1))v^2\bigg)\Bigg).
\end{align*}
We can then find the remaining component size from $S=p_s-p_sg(1-\pi+\pi u,(1-\pi)^2+2(1-\pi)^2\pi v +\pi^2(3-2\pi)v^2).$

\end{document}